\documentclass[aps,epsfig,twocolumn]{revtex4}

\usepackage{graphicx}
\usepackage{dcolumn}
\usepackage{bm}

\newcommand{\bq}{\begin{equation}}
\newcommand{\eq}{\end{equation}}
\newcommand{\bqa}{\begin{eqnarray}}
\newcommand{\eqa}{\end{eqnarray}}
\newcommand{\nn}{\nonumber \\}

\begin{document}
\draft
\title{ 
U(1) Gauge Theory of the Hubbard Model :
Spin Liquid States and Possible Application to
$\kappa-(BEDT-TTF)_2 Cu_2 (CN)_3$ 
}

\author{Sung-Sik Lee and Patrick A. Lee}
\address{Department of Physics, Massachusetts Institute of Technology,\\
Cambridge, Massachusetts 02139, U.S.A.\\}
\date{\today}
       
\begin{abstract}
We formulate a U(1) gauge theory of the Hubbard model in the slave-rotor representation.
From this formalism it is argued that spin liquid phases may exist near the Mott transition in the Hubbard model on triangular and honeycomb lattices at half filling.
The organic compound $\kappa-(BEDT-TTF)_2 Cu_2 (CN)_3$ is a good candidate for the spin liquid state on a triangular lattice.
We predict a highly unusual temperature dependence for the thermal conductivity of this material.
\end{abstract}
\maketitle

\newpage
The Hubbard Hamiltonian is the simplest model which exhibits the Mott transition.
For small Coulomb repulsion the Fermi liquid phase occurs.
In the large Coulomb repulsion limit the model reduces to the t-J model which has insulating ground state and spin order at half filling. 
In the intermediate region a spin liquid phase, i.e. an insulator without spin order, may arise.
This spin liquid phase is more likely to occur if there is frustration.

Recent experiments indicate that the spin liquid ground state may be realized in the organic compound $\kappa-(BEDT-TTF)_2 Cu_2 (CN)_3$ which is just on insulating side of the Mott transition\cite{SHIMIZU}.
While the high temperature spin susceptibility can be fitted to a Heisenberg $S=1/2$ model with an exchange energy of $250K$, the spins do not order down to $32 mK$.
The susceptibility and Knight shift\cite{KAWAMOTO} drop sharply below $\sim 20K$, but remains finite down to the lowest temperature.
The Carbon nuclear spin relaxation rate $1/T_1T$ also saturates to a finite value at low $T$.
An interpretation in terms of a spin liquid on a frustrated lattice has been proposed\cite{SHIMIZU} but an alternative explanation in terms of localization has also been suggested\cite{KAWAMOTO}.
This system is effectively described by a Hubbard model at half filling on a 2-dimensional triangular lattice.
On this lattice the Heisenberg model is known to have an antiferromagnetic (AF) ground state.
Recently Motrunich has shown using projected trial wavefunctions that a spin liquid state with spinon Fermi surface may be stable if the t-J model is extended to include higher order virtual hoppings, such as the ring exchange terms\cite{MOTRUNICH}.
If true, this will be a realization of the resonating valence bond (RVB) idea of Anderson\cite{ANDERSON}.
In order to include such charge fluctuations effects, it is clearly desirable to study the Hubbard model near the Mott transition.

There are numerical evidences that the Hubbard model on the triangular lattice does not have AF order for moderate Coulomb repulsion in the insulating phase\cite{IMADA}.
However, the nature of the disordered state is not understood.
Recently Florens and Georges\cite{FLORENS} introduced the slave-rotor representation of the Hubbard model.
This method is more economical than the conventional representation\cite{KOTLIAR} which requires $4$ slave bosons.
From this theory the Mott transition on the square lattice was successfully described at the mean-field level\cite{FLORENS}. 
Thus it is of great interest to see whether the spin liquid phase is predicted by the slave-rotor theory in the triangular lattice.
If a spin liquid mean-field state is found, the next question is the stability of the state.
Although it is expected that the U(1) gauge theory will emerge as a low energy theory, it is not clear how to derive it from the mean-field theory of Ref. \cite{FLORENS}.
The first objective of the present paper is to formulate a U(1) gauge theory of the Hubbard model in the slave-rotor representation.
Then we apply the formalism to the triangular and honeycomb lattices to find the spin liquid mean-field states.
The low energy theories of some spin liquid states are shown to be a compact U(1) gauge theory coupled with gapless spinons. 
Finally we discuss the deconfinement phase of the U(1) gauge theory in connection with the experimentally observed spin liquid behavior in the organic material.

The Hubbard Hamiltonian reads
\bq
H  = 
-\sum_{<i,j>,\sigma}  ( t_{ij}   c_{i\sigma}^\dagger c_{j\sigma} + h.c. )
+ \frac{U}{2} \sum_i \left( \sum_{\sigma} c_{i\sigma}^\dagger c_{i\sigma} - 1 \right)^2.
\eq
Here $c_{i \sigma}$ is the annihilation operator of electron with spin $\sigma$ at site $i$.
$t_{ij} = |t_{ij}| e^{-i A_{ij}}$ is the complex hopping integral with $A_{ij}$ representing either external electromagnetic (EM) vector potential or an intrinsic phase coming from the overlap of atomic wavefunctions.
$U$ is the on-site Coulomb repulsion.
In the slave-rotor representation the electron operator is written as
$c_{i \sigma} = f_{i \sigma}  e^{-i \theta_i}$,
 where $f_{i \sigma}$ is the spinon annihilation operator 
and $e^{-i \theta_i}$, the lowering operator of the `angular momentum' $L_i$ which corresponds to the charge quantum number.
The enlarged Hilbert space is constrained by $L_i = \sum_{\sigma} f_{i \sigma}^\dagger f_{i \sigma} - 1$.
The partition function is written as a path integral of $e^{-S_0}$ over $f$, $f^*$, $\theta$ and $h$ where
\bqa
S_0
 =  \int_0^\beta d \tau \Bigl[ 
\sum_{i,\sigma} f_{i \sigma}^*  ( \partial_\tau + ih_i - \mu ) f_{i \sigma} 
 + \frac{1}{2U} \sum_i ( \partial_\tau \theta_i + h_i )^2  && \nn
- \sum_{<i,j>}  ( t_{ij} f_{i\sigma}^* f_{j\sigma} e^{i (\theta_i - \theta_j)} + c.c. ) 
 + \sum_i \left(  \mu (1-x) - i h_i \right) \Bigr]. &&
\label{s1}
\eqa
Here $h_i$ is the Lagrangian multiplier field imposing the constraints
$L_i - ( \sum_{\sigma} f_{i \sigma}^\dagger f_{i \sigma} - 1 ) = 0$,
$\mu$ is the chemical potential and $x$, the hole doping density from half filling.
In the above action every term is quadratic in $f$ and $\theta$ except the hopping term.
In Ref.\cite{FLORENS} the hopping term was decomposed as 
$\alpha \beta \approx <\alpha> \beta + \beta <\alpha> - <\alpha><\beta>$ with 
$\alpha = \sum_\sigma f_{i\sigma}^* f_{j\sigma}$ and 
$\beta = e^{i (\theta_i - \theta_j)}$. 
Here we devise a more systematic way of decomposing the hopping term, so that the saddle point coincides with the previous mean-field theory.
The merit of the present method is that it enables us to expand about the saddle point to obtain a low energy effective theory. 
To decompose the hopping term we use the identity
\bqa
&& e^{\epsilon ( \alpha_{ij} \beta_{ij} + \alpha_{ji} \beta_{ji} ) }
 = \frac{\epsilon^2}{\pi^2 } \int d\eta_{ij} d\eta_{ij}^* d\eta_{ji} d\eta_{ji}^* \nn
&& e^{ -\epsilon \left[
  |\eta_{ij}|^2 + |\eta_{ji}|^2
 - \alpha_{ij} \eta_{ij} - \alpha_{ji} \eta_{ji}
 - \beta_{ij} \eta_{ij}^* - \beta_{ji} \eta_{ji}^*
\right] }
\label{ab}
\eqa
and set
$\epsilon  =  \Delta \tau |t_{ij}|$, 
$\alpha_{ij}  =  \sum_{\sigma} f_{i\sigma}^* f_{j\sigma}$ and
$\beta_{ij}  =  e^{i (\theta_i - \theta_j - A_{ij})}$.
This is a bit different from the conventional Hubbard-Stratonovich decomposition of a complete square and we encounter two problems.
First, we need to introduce two independent complex fields $\eta_{ij}$ and  $\eta_{ji}$ to decompose the hopping term and its complex conjugate, so that for a given configuration of $\eta_{ij}$ and  $\eta_{ji}$, the action is not real.
Second, it appears that the same $\eta_{ij}$ field decouples the fermion and the rotor term and it is not clear how we can recover the Florens-Georges's mean-field theory where $<\alpha> \neq <\beta>$.
Both difficulties are resolved as follows.
We change the variables of integration by 
$\eta_{ij}  =  |\chi_{ij}| e^{w_{ij}  + i (a^+_{ij} + a_{ij}) }$ and
$\eta_{ji}  =  |\chi_{ij}| e^{-w_{ij} + i (a^+_{ij} - a_{ij}) }$.
Here $|\chi_{ij}|$ parameterizes the geometric mean of $|\eta_{ij}|$ and $|\eta_{ji}|$,
and $w_{ij}$, the relative amplitudes.
$a_{ij}^+$ and $a_{ij}$ are respectively the symmetric and antisymmetric combinations of the phases of $\eta_{ij}$ and $\eta_{ji}$. 
It is emphasized that $w_{ij}$ and $a_{ij}^+$ are necessary in the integration because $\eta_{ij}$ and $\eta_{ji}$ are independent complex variables (not complex conjugate to each other). 
We follow Florens and Georges to replace $e^{i \theta_i}$ by the boson variable $X_i$ with a constraint
$|X_i| = 1$ which is imposed by a Lagrangian multiplier $\lambda_i$.
The partition function now takes the form
\bqa
Z & = &
\int D |\chi_{ij}| Dw_{ij} Da_{ij} Da_{ij}^{+}
D f D f^* D X D X^* Dh D\lambda \nn
&& \left[ \Pi_{<ij>,\tau} (4 |\chi_{ij}|^3) \right]
e^{-S},
\label{z1}
\eqa
where
\bqa
&& S
 =  \int_0^\beta d \tau \Bigl[
2 \sum_{<i,j>}  |t_{ij}|  |\chi_{ij}|^2 \cosh {2 w_{ij}} \nn
&& + \sum_i \left(  \mu (1-x) - i h_i - i \lambda_i \right)  
 + \sum_{i,\sigma} f_{i \sigma}^*  ( \partial_\tau + ih_i - \mu ) f_{i \sigma} \nn
&& - \sum_{<i,j>,\sigma}  |t_{ij}|  |\chi_{ij}| e^{ia_{ij}^+} \left(
          e^{w_{ij} + i a_{ij} }  f_{i\sigma}^* f_{j\sigma}
       +  e^{ -w_{ij} - i a_{ij} } f_{j\sigma}^* f_{i\sigma}
     \right) \nn
&& +  \frac{1}{2U} \sum_i  [ ( i \partial_\tau  + h_i ) X_i^*] [( -i \partial_\tau  + h_i ) X_i ]
 + i  \sum_i \lambda_i |X_i|^2  \nn
&& - \sum_{<i,j>}  |t_{ij}|  |\chi_{ij}| e^{-ia_{ij}^+} \Bigl(
           e^{w_{ij} - i a_{ij} -i A_{ij} }  X_j^* X_i  \nn
&&        +  e^{ -w_{ij} + i a_{ij} + i A_{ij} }  X_i^* X_j
     \Bigr) \Bigr].
\label{s2}
\eqa
%\end{widetext}
Note that $a_{ij}$ and $h_i$ correspond to the gauge field associated with the U(1) gauge transformation :
$f_{i \sigma} \rightarrow e^{i \varphi_i} f_{i \sigma}$,
$X_{i } \rightarrow e^{i \varphi_i} X_{i}$,
$a_{ij} \rightarrow a_{ij} + ( \varphi_i - \varphi_j)$ and
$h_{i } \rightarrow h_{i} - \partial_\tau \varphi_i$.
Here we choose to couple the external EM field ($A_{ij}$) to the boson.
Alternatively $A_{ij}$ can be coupled to spinon by a shift of the gauge field $a_{ij} \rightarrow a_{ij} - A_{ij}$.

The action (\ref{s2}) remains real for the fluctuations of $|\chi_{ij}|$ and $a_{ij}$.
It becomes complex for the fluctuations of $w_{ij}$, $a^+_{ij}$, $\lambda_i$ and $h_i$.
For the latter set of variables we lift the contour of integration into complex plane via analytic continuation to find a saddle point in the imaginary axis.
Allowing for fluctuations along the deformed contour we write
$|\chi_{ij}|  =  |\tilde \chi_{ij}| + \delta |\chi_{ij}|$,
$a_{ij}  =  \tilde a_{ij} + \delta a_{ij}$,
$w_{ij}  =  i \tilde w_{ij} + \delta w_{ij}$, 
$a_{ij}^+  =  i \tilde a^+_{ij} + \delta a^+_{ij}$, 
$\lambda_i  =  i \tilde \lambda_i + \delta \lambda_i$ and
$h_i  =  i \tilde h_i + \delta h_i $,
where quantities with tildes are real and represent the saddle point value. 
The appearance of imaginary saddle point is familiar in the treatment of the constraint field in the usual slave-boson models.
In particular, fluctuations in the $\delta a^+_{ij}$ and $\delta w_{ij}$ directions are stable and massive.
Together with $\delta |\chi_{ij}|$, these massive modes are negligible for energy scale less than $t$ or $U$.
The low energy effective Lagrangian is reduced to fermions and $X$ bosons coupled to the compact U(1) gauge fields $a_{ij}$ and $h_i$, together with a constraint field $\lambda_i$.
If we choose, we can integrate over the $\lambda_i$ field to restore the $\theta$ field resulting in the effective Lagrangian
\bqa
&& L^{'} = 
 \sum_{j,\sigma} f_{j \sigma}^*  ( \partial_\tau - i a^\tau_j  - \tilde h_i - \mu ) f_{j \sigma} \nn
&& + \frac{1}{2U} \sum_i   (  \partial_\tau \theta_i  - a^\tau_i + i \tilde h_i )^2 
- \sum_{j,\alpha,\sigma}  |t_{\alpha}| \tilde \chi_{\alpha}^X  e^{i a^{\alpha}_j} f_{j+\alpha \sigma}^* f_{j\sigma} 
\nn
&&  - \sum_{j,\alpha}  |t_\alpha| \tilde \chi_\alpha^f e^{-i ( \theta_{j+\alpha} - \theta_j -  a^{\alpha}_j - A^\alpha_j )},
\label{s3}
\eqa
where $\alpha$ refers to the direction of neighboring sites connected by the hoppings.
$a^{\alpha}_j$ and $a^\tau_j$ are respectively the spatial and temporal gauge fields coming from $\delta a_{j+\alpha, j}$ and $\delta h_j$ in (\ref{s2}).
The saddle point ($\tilde \chi_{ij}^f$, $\tilde \chi_{ij}^X$, $\tilde h_i$, $\tilde \lambda_i$) is determined from the extremum condition of the free energy and is identical to the mean-field theory of Florens and Georges\cite{FLORENS},
where
$\tilde \chi_{ij}^f  =  |\tilde \chi_{ij}| e^{\tilde a^+_{ij} - i (\tilde w_{ij}-\tilde a_{ij})}$ and
$\tilde \chi_{ij}^X  =  |\tilde \chi_{ij}| e^{- \tilde a^+_{ij} + i (\tilde w_{ij}+\tilde a_{ij})}$.
It is interesting to note that the auxiliary fields associated with the amplitude ($w_{ij}$) and phase ($a^+_{ij}$) fluctuations in the original integration (\ref{z1}) switched their roles to make the free energy real at the saddle point and resolve both difficulties mentioned earlier.
Eq. (\ref{s3}) is the first main result of this paper.

Here we remark on the connection between the slave-rotor theory\cite{FLORENS} and the conventional slave-boson theory\cite{KOTLIAR}.
In the slave-rotor theory the boson is relativistic while the conventional slave-boson fields are non-relativistic.
The relativistic $X$ boson contains particle and anti-particle. 
In the real-time canonical quantization the boson operator is written
$X(x,t) \sim \int \frac{d k^2}{ E_k} 
\left(   
  b_k e^{-i ( E_k t - k x)}
+ d_k^\dagger e^{i ( E_k t - k x)}
\right)$,
where the $b_k$ ($d_k$) is the (anti) particle annihilation operator
and $E_k$ is the boson energy dispersion. 
$b_k$ ($d_k$) carries positive (negative) charges for both external ($A_{ij}$) and internal ($a_{ij}$) gauge fields.
Thus we identify $b_k^\dagger$ as creating holon and $d_k^\dagger$ as doublon.
In the insulating phase the bosons are gapped, and holon and doublon are bound.
In the metallic phase they becomes gapless with unbound holon and doublon.

We apply the above formalism to find mean-field states in the triangular and honeycomb lattices at zero temperature.
We focus on the half filling where $\tilde h = 0$.
For the triangular lattice we consider the flux phase :
$\chi_{ij}^f = \chi^f e^{i \phi^f_{ij}}$  and $\chi_{ij}^X = \chi^X e^{i \phi^X_{ij}}$, 
where the phases are arranged so that the spinon (boson) acquires a flux when it moves around the smallest triangular plaquette.
Here we choose $t>0$, which can be always made by a particle-hole transformation at half-filling.  
We have checked that the staggered flux states, where the fluxes take on alternating signs, are alway unstable relative to the zero flux state in the parameter range studied.
For the Heisenberg model on the triangular lattice the fully gapped state with $\pi$-flux through the unit cell (henceforth, `$\pi$-flux phase') is known to be stable among paramagnetic states\cite{TKLEE}.
This is a state with flux $\pi/2$ through each triangle; 
it breaks time reversal symmetry and is closely related to the chiral spin state\cite{KALMEYER}.
It is of interest to see whether a gapless state without flux (henceforth, `uniform phase') can be stabilized by charge fluctuations near the Mott transition.
In our mean field theory the metallic phase is characterized by a nonvanishing amplitude of bose condensation which is given by 
$Z = \frac{1}{N}< X^*(k=0, \tau) X(k=0, \tau) >$, where $X(k, \tau)$ is the boson field in the momentum space and $N$, the number of unit cells.
For the isotropic triangular lattice a first order transition to the $\pi$-flux phase is found to occur at $U_c/t \approx 2.7$ as is shown in Fig. 1.
From finite size scaling we found that the uniform phase is stable over $2.6 < U/t < 2.73$ as is shown in the inset of Fig. 1.
This insulating phase corresponds to the spin liquid state with a Fermi surface of neutral spinon.
In order to see how the spin liquid phase is affected by anisotropy in the hopping integral we consider $t^{'} \neq t$ with $t^{'}$, the hopping integral in one anisotropic direction.
As $|t^{'}-t|/t$ increases the region of the spin liquid phase shrinks and closes at $|t^{'}-t|/t \approx 0.1$.
In comparison with the numerical result\cite{IMADA} there are quantitative discrepancies. 
For example, Imada et al.\cite{IMADA} found $U_c/t \approx 5$ for $t^{'}=t$ and the spin liquid state is found to be gapless and therefore not the chiral state already for $t^{'}/t = 0.5$. 
The mean field phase diagram is not expected to be quantitative and it is apparent that the stability of the $\pi$-flux phase has been overestimated.
Rather, the usefulness of our approach lies in obtaining a description of the insulating state and we now proceed to analyse the uniform phase.

In the insulating phase the rotor has a gap and can be integrated out.
This generates the Maxwellian kinetic term, 
${\cal L } = \frac{1}{2 g^2} (\partial \times a)^2$ 
with the gauge coupling,
$g^2 \sim \frac{\Delta^2}{\kappa}$ with
$\Delta$, the charge gap and     
$\kappa \sim t \tilde \chi^f$, the bare phase stiffness of boson\cite{NAGAOSA2000}.
The low energy theory is reduced to the compact U(1) gauge theory coupled with spinons with Fermi surface. 
The compact U(1) pure gauge theory is always confining\cite{POLYAKOV} but whether deconfinement is possible in the presence of matter field is an open question.
Herbut et al. has argued that the theory is always confining in the presence of Fermi surface\cite{HERBUT_FS} or nodal fermions\cite{HERBUT_NP}.
Their conclusions actually apply to an approximate effective action for the gauge field obtained by integrating out the fermions.
Integrating out the fermions is suspect in the presence of gapless fermions.
Indeed recently Hermele et al.\cite{HERMELE} proved that if the spin index is generalized to N flavors,
the problem of 2N 2-component Dirac fermions coupled to complex U(1) gauge fields is deconfined for sufficiently large N, thus providing a counter example to Ref.\cite{HERBUT_NP}.
The spinon Fermi surface contains even more low energy excitations.
In our view it is likely that a deconfinement state is possible and we will proceed with the assumption.

There is extensive work on the coupling of the Fermi sea with non-compact gauge fields.
A controlled perturbation in $1/N$ expansion is possible\cite{IOFFE,PLEE92,POLCHINSKY}.
For small $q$ and $\omega$, a quantum Boltzmann description is possible, albeit with singular Landau parameters which lead to anomalous power law behavior of many physical quantities\cite{KIM95}.
For $q = 2k_F$, there are singularities in the vertex function and it remains open whether the spin susceptibility is divergent or not\cite{ALTSHULER}.
In contrast to the square lattice, there is no Fermi surface nesting for the half filled triangular lattice and antiferromagnetic order does not emerge naturally in our mean-field theory.
Thus we believe that the spinon Fermi surface is stable and is the likely candidate for the spin liquid state found numerically\cite{IMADA} as well as experimentally\cite{SHIMIZU}.
Our conclusion is in agreement with that of Motrunich\cite{MOTRUNICH}.
Of course, there remain the possibility of pairing instability due to residual interaction in any Fermi liquid, but that may set in at a very low energy scale.

What are the physical implications of the spinon Fermi surface?
Low lying gauge fluctuations produce a $T^{2/3}$ contribution to the specific heat\cite{MOTRUNICH}.
The spinons carry spin and entropy.
The spin conductivity (which is difficult to measure) is proportional to the spinon transport time due to scattering by gauge fluctuations which goes as $T^{-4/3}$\cite{PLEE92}.
On the other hand, the thermal conductivity is proportional to the energy relaxation time which goes as $T^{-2/3}$ because small $q$ gauge fluctuations are effective in relaxing the nonequilibrium energy distribution.
Using the quantum Boltzmann description\cite{KIM95} we predict that $\kappa /T \sim T^{-2/3}$ in the clean limit.
The divergent behavior will be cut off by impurity scattering.
Of course, being an insulator the charge conductivity is zero.
This gross violation of the usual Wiedermann-Franz law together with unusual power laws are unique signatures of this particular spin liquid state.
In view of the unusual behavior of $\kappa-(BEDT-TTF)_2 Cu_2 (CN)_3$ it will clearly be of great interest to perform specific heat and thermal conductivity measurements in these materials.

We have also applied the mean field theory to the honeycomb lattice.
There is no obvious experimental candidate but the interest in the honeycomb lattice is mainly theoretical.
For nearest neighbor hopping, the fermion dispersion is characterized by 2 inequivalent Dirac cones at the Brillouin zone corner.
On the insulating side of the Mott transition, the rotor is again gapped and the problem reduces to 2N Dirac cones coupled to a compact U(1) gauge field with $N=2$.
As mentioned earlier, there is hope that the theory is deconfined\cite{HERMELE}.
Furthermore, the linear density of states greatly reduces the spin susceptibility at a wavevector connecting the modes, and the AF ordered state does not emerge naturally in our mean field theory.
It is amusing to observe that the energy per bond of the naive AF state ($-J/4$) is identical to that of the dimer trial wavefunction.
In this sense the honeycomb lattice is between the one dimensional chain and the square lattice.
Our hope is that the AF state is not the ground state near the Mott transition, even though we can not address the competition directly in our mean field theory.
Then the remaining competition is between the dimer state (which breaks rotational symmetry) and the nodal spin liquid.
Although not shown here we find for the nearest neighbor Hubbard model the nodal spin liquid state is stable over the narrow range $1.68 < U/t < 1.74$ near the Mott transition.
Here we add further hopping terms $t^{'}$ and $t^{''}$ as shown in Fig. 2. 
It is straightforward to show that $t^{'}$ does not affect the phase diagram or the energy as long as it is small compared to $t$.
For negative $t^{''}/t$, we found that the spin liquid state is stable over a wider range of $U/t$, e.g., $2.3 < U/t < 2.6$ for $t^{''}/t = -0.4$ as is shown in Fig. 2. 
This region is the spin liquid phase with nodal spinons.
Unusual power law for the spin response function corresponding to AF fluctuations has been proposed for the state\cite{RANTNER}.
Unlike the triangular lattice, the Hubbard model on the honeycomb lattice with $t$ and $t^{''}$ can be studied by quantum Monte Carlo without the Fermi sign problem so that a definite answer should be attainable\cite{SORELLA,MONTHOUX}.

We thank T. Senthil, A. Georges and X.-G. Wen for many illuminating discussions.
This work was supported by the Post-doctoral Fellowship Program of Korea Science $\&$ Engineering Foundation (KOSEF) and NSF grant DMR-0201069.

%\references

\begin{figure}
        \includegraphics[height=5cm,width=8cm]{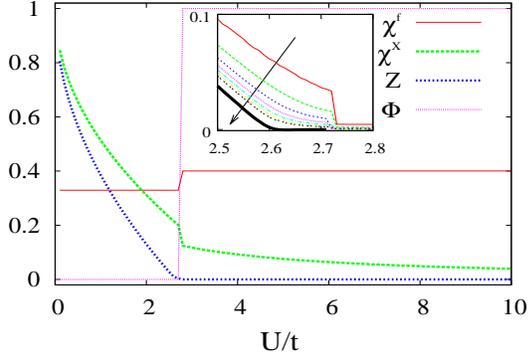}
\label{fig:1}
\caption{
The amplitudes of hopping order parameters of spinon ($\chi^f$) and boson ($\chi^X$), the bose condensation amplitude($Z$) and the uniform flux ($\Phi$ in unit of $\pi$) per unit cell in the isotropic triangular lattice at half filling and $T=0K$. 
$50 \times 50$ lattice is used for the figures in the main pannel.
The inset shows the size dependence of $Z$ near the Mott transition.
The arrow indicates the direction of increasing size from $N= 20 \times 20$ to $80 \times 80$.
The thick line denotes $Z$ in the thermodyamics limit obtained from the finite size scaling.
}
\end{figure}

\begin{figure}
        \includegraphics[height=5cm,width=8cm]{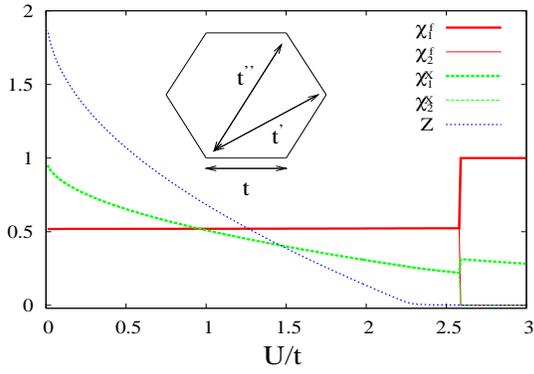}
\label{fig:2}
\caption{
The amplitudes of the nearest neighbor hopping order parameters of spinon ($\chi^f$) and boson($\chi^X$), and the bose condensation amplitude($Z$) in the honeycomb lattice with $t^{'}/t = 0$ and $t^{''}/t = -0.4$ at half-filling and $T=0K$.  The subscripts $1$ and $2$ refers to two inequivalent bonds in the dimerized phase.
}
\end{figure}
%\end{widetext}

\end{document}